\documentclass[]{arxsub}

\usepackage{ifluatex}
\ifluatex
\usepackage{fontspec}
\usepackage{polyglossia}
\setdefaultlanguage{english}
\else
\usepackage[english]{babel}
\usepackage[utf8]{inputenc}
\fi

\usepackage[
  backend=biber,
  natbib = true,
  style=ext-numeric-comp,
  sorting=none,
  minbibnames=3, maxbibnames=6,
  uniquename=false, uniquelist=false,
  giveninits=true, terseinits,
  articlein=false, innamebeforetitle=true,
  isbn=false,
  urldate=long, dateabbrev=false,
  autocite=superscript,
]{biblatex}

\DeclareNameAlias{sortname}{family-given}

\DeclareNameAlias{author}{sortname}
\DeclareNameAlias{editor}{sortname}
\DeclareNameAlias{translator}{sortname}

\DeclareDelimAlias{finalnamedelim}{multinamedelim}

\DeclareFieldFormat
  [article,inbook,incollection,inproceedings,patent,thesis,unpublished]
  {title}{#1}

\renewbibmacro*{journal+issuetitle}{\usebibmacro{journal}\setunit*{\addspace}\iffieldundef{series}
    {}
    {\newunit
     \printfield{series}\setunit{\addspace}}\usebibmacro{issue+date}\setunit{\addsemicolon\addspace}\usebibmacro{volume+number+eid}\setunit{\addcolon\space}\usebibmacro{issue}}

\DeclareFieldFormat[article,periodical]{number}{\mkbibparens{#1}}

\renewbibmacro*{issue+date}{\usebibmacro{date}}

\renewbibmacro*{pubinstorg+location+date}[1]{\printlist{location}\iflistundef{#1}
    {\setunit*{\addsemicolon\space}}
    {\setunit*{\addcolon\space}}\printlist{#1}\setunit*{\addsemicolon\space}\usebibmacro{date}\newunit}

\DeclareFieldFormat{pages}{#1}

\DeclareFieldFormat{doi}{doi\addcolon\space
  \ifhyperref
    {\href{https://doi.org/#1}{\nolinkurl{#1}}}
    {\nolinkurl{#1}}}

\AtEveryBibitem{\clearfield{number}\clearfield{month}\clearfield{day}\clearfield{urlyear}\clearfield{urlmonth}\clearfield{issn}\clearfield{series}\clearfield{pagetotal}\clearfield{eprintclass}\ifentrytype{article}{
	\iffieldundef{volume}{}{\clearfield{doi}}\clearfield{eprint}\clearfield{url}}{}\ifentrytype{book}{\clearfield{eprint}\clearfield{doi}\clearfield{url}}{\clearname{editor}}
\ifentrytype{inproceedings}{\clearfield{eprint}\clearfield{url}\clearfield{doi}\clearfield{publisher}\iffieldundef{booktitle}{}{\clearfield{eventtitle}}}{}\iffieldundef{eprint}{}{\clearfield{url}}
\iffieldundef{doi}{}{\clearfield{url}}
}

\usepackage{csquotes}
\usepackage{siunitx}
\DeclareSIUnit\pixel{pixel}
\usepackage[section]{placeins}

\usepackage{hyperref}
\usepackage[capitalise,noabbrev]{cleveref} 

\newcommand{\ie}{i.\,e.\ }
\newcommand{\eg}{e.\,g.\ }
\newcommand{\cf}{cf.\ }
\usepackage{listings}
\usepackage{booktabs}

\providecommand{\doi}{}
\renewcommand{\doi}[1]{\href{https://doi.org/#1}{#1}}

\newcommand{\preSubmission}{Part of this work has been presented at the ISMRM Annual Conference 2025 (Hawaii).}

\providecommand{\del}[2][]{}
\providecommand{\dels}[2][]{}

\makeatletter
\appto{\appendix}{\setcounter{figure}{0}}
\makeatother

\begin{document}

\articletype{Toolbox Article}
\title{BART Streams: Real-time Reconstruction
Using a Modular Framework for Pipeline Processing}

\author[1,2]{Philip Schaten}
\author[1,3]{Moritz Blumenthal}
\author[1]{Bernhard Rapp}
\author[2,4,5]{\\Christina Unterberg-Buchwald}
\author[1,2,4,6]{Martin Uecker}

\authormark{Schaten \textsc{et al}}

\address[1]{Institute of Biomedical Imaging, Graz University of Technology, Graz, Austria}
\address[2]{Institute for Diagnostic and Interventional Radiology, University Medical Center Göttingen, Germany}
\address[3]{Department of Radiology, Boston Children's Hospital, Harvard Medical School, Boston, USA}
\address[4]{German Centre for Cardiovascular Research (DZHK), partner site Lower Saxony, Germany}
\address[5]{Clinic for Cardiology and Pneumology, University Medical Center Göttingen, Germany}
\address[6]{BioTechMed-Graz, Graz, Austria}

\corres{Martin Uecker,
Graz University of Technology,
Institute of Biomedical Imaging,\\
Stremayrgasse 16/3,
8010 Graz,
AUSTRIA,
\email{uecker@tugraz.at}}

\finfo{This work was funded by DZHK (German Centre for Cardiovascular Research) grant 81X4300119 and
in part by NIH under grant U24EB029240.}

\abstract{
\section{Purpose}
To create modular solutions for interactive real-time MRI using reconstruction algorithms implemented in BART.
\section{Methods}
A new protocol for streaming of multidimensional arrays is presented and integrated into BART.
The new functionality is demonstrated using examples for cardiac interactive real-time MRI
based on radial FLASH, where iterative reconstruction is combined with advanced 
features such as dynamic coil compression and gradient-delay correction.
We analyze the latency of the reconstruction and measure end-to-end latency of
the full imaging process.
\section{Results}
Reconstruction pipelines with iterative reconstruction and advanced functionality
were built in a modular way using scripting.
Latency measurements demonstrate latency sufficient for interactive real-time MRI, on the order of \qty{30}{ms} for BART processing
and network transfer time, or \qty{200}{ms} for end-to-end latency including acquisition, vendor processing, and display.
\section{Conclusion}
With the new streaming capabilities, real-time reconstruction pipelines can be 
assembled using BART in a flexible way, enabling rapid prototyping of advanced 
applications such as interactive real-time MRI.
}


\jnlcitation{\cname{\author{P. Schaten},
\author{M. Blumenthal},
\author{M. Uecker}} (\cyear{2025}),
\ctitle{Real-Time MRI Reconstruction with BART Streams},
\cjournal{Magn. Reson. Med.},
\cvol{???}.}

\maketitle

\section{Introduction}

\enquote{Live streams} from inside the human body
were possible for a long time \cite{Riederer_Magn.Reson.Med._1988}
through the combination of real-time MRI acquisition,
real-time reconstruction and low-latency visualization.
We refer to this as interactive real-time MRI (RT-MRI)
\cite{Nayak_Magn.Reson.Med._2019,Dietz_MagnResonMed_2019}.
It is an important tool with emerging clinical applications, such as interventional
\cite{Buecker_J.Magn.Reson.Im._2000,Ratnayaka_Eur.HeartJ._2013,Campbell-Washburn_Magn.Reson.ImagingClin.N._2015,Unterberg-Buchwald_J.Cardiov.Magn.Reson._2017,Nageotte_Pediatr.Cardiol_2020} 
or fetal MRI \cite{Silva_Magn.Reson.Med_2023}.
RT-MRI usually requires high frame rates,
\eg to resolve cardiac motion \cite{Nayak_Magn.Reson.Med._2001},
speech production \cite{Narayanan_J.Acoust.Soc.Am._2004,Niebergall_Magn.Reson.Med._2013},
or joint motion \cite{Rasche_Magn.Reson.Med._1995}.
Thus, RT-MRI often makes use of data undersampling
in combination with specific reconstruction methods,
for example view-sharing \cite{Riederer_Magn.Reson.Med._1988},
parallel imaging \cite{Saybasili_Magn.Reson.Imaging_2014},
compressed sensing \cite{Lustig_Magn.Reson.Med._2007,Majumdar_IEEETrans.Med.Imag._2012},
Kalman filtering \cite{Sumbul_IEETrans.Med.Imag._2009_2},
spatiotemporally constrained reconstruction \cite{Le_Magn.Reson.Med._2025},
or other.

In this work, we show examples using 
the real-time regularized non-linear inversion (RT-NLINV) algorithm
\cite{Uecker_NMRBiomed._2010,Uecker_Magn.Reson.Med._2010}.
Based on classic regularized non-linear inversion (NLINV)
\cite{Uecker_Magn.Reson.Med._2008},
RT-NLINV offers particularly high undersampling factors
and is causal, \ie it does not rely
on data acquired after the frame that is currently being reconstructed.
Running iterative RT-MRI reconstruction algorithms such as RT-NLINV
in real-time typically requires
optimized software and hardware acceleration
\eg through the use of Graphical Processing Units (GPU) 
 \cite{Nayak_JMR_2022,Sorensen_IEEETrans.Med.Imag._2009,Schaetz__2012}, especially when low latency is required.
For example, cardiac interventional MRI and MRI-guided
radiotherapy might require a latency as low as several \qty{100}{ms}
\cite{Unterberg-Buchwald_J.Cardiov.Magn.Reson._2017,Green_Med.Phys._2018},
and speech biofeedback studies were performed with a latency 
of \qty{100}{ms} \cite{Kumar__2024}.

The BART toolbox for computational MRI \cite{bart__2023} is a comprehensive framework which
provides a wide range of reconstruction,
calibration \cite{Uecker_Magn.Reson.Med._2014},
machine-learning \cite{Blumenthal_Magn.Reson.Med._2023,Blumenthal_Magn.Reson.Med_2024},
and signal processing tools.
Its optimized code and its existing support for GPU computing make it an ideal choice for
the development of real-time reconstruction methods.
Another core feature of BART is its modular design as a toolbox: Algorithms are made available
as individual command line tools,
enabling the construction of powerful reconstruction pipeline by
combining the different tools in a script.
However, high-quality, low latency RT-MRI requires not only
 optimized reconstruction, but also the
ability to stream the data and reconstructed
images during the measurement.
Each tool in the reconstruction pipeline
needs to perform its task as soon as a raw-data frame
is available, as opposed to waiting for a complete dataset.
Existing software such as the Gadgetron \cite{Hansen_Magn.Reson.Med._2013},
can stream data in ISMRMRD format \cite{Inati_Magn.Reson.Med_2016}, and
BART has  been used with Gadgetron in the past
\cite{Diakite__2018,Veldmann_Magn.Reson.Med_2022}.
However, up to now, it was not possible to use BART
to construct low-latency processing pipelines.

In this work, we integrate new stream processing capabilities into BART.
The new functionality facilitates building
reconstruction pipelines suitable for RT-MRI
with existing BART tools.
Importantly, our approach to streaming
builds on standard Unix system utilities, which allows us to preserve the full
modularity of BART. This is demonstrated using examples for interactive real-time MRI
based on radial FLASH, where iterative reconstruction is combined with advanced 
features such as dynamic coil compression and gradient-delay correction.
In particular, we show how the RT-NLINV algorithm implemented in
BART can now be used in real-time reconstruction pipelines, and how a
streaming version of geometric coil compression \cite{Zhang_Magn.Reson.Med._2013}
acquires a new use as dynamic coil compression for interactive RT-MRI.
Additionally, we analyze the latency of the
reconstruction itself and measure the
end-to-end latency of the complete imaging process.

\section{Methods}

We will first give an overview over the architecture of BART, then discuss
multi-dimensional arrays, and then explain the new support for looping,
and streaming, which is then used for real-time processing.

\subsection{BART Structure}

BART is a modular reconstruction framework which enables building complex reconstruction
algorithms through combination of various individual components.
On the lowest level, a numeric library provides generic functions for operations
on multidimensional arrays (\textit{md-arrays}).
Many of these operations can be accelerated using GPUs,
or with the Message Passing Interface (MPI) \cite{mpi41} toolkit,
enabling distributed computing on scientific high-performance clusters
and multi-GPU usage \cite{Blumenthal_ISMRM_2025}.
The multi-dimensional operations are complemented by an operator library,
which provides a simple way
of constructing operators, \eg the forward operator describing an MRI measurement, while automatically
providing its adjoint in the case of a linear operator, or the derivative for non-linear
operators.

For most use cases it is sufficient to interact 
with BART using its command line interface, which consists of a set of high-level
tools operating on md-arrays.
All commands take several individual options and typically input and output file names.
A single driver command, \texttt{bart},  acts as the
unified entry point to BARTs command line interface.
It can be called directly from a command line or from scripting languages 
such as BASH, Python or Matlab.
The \texttt{bart} call is followed by a set of general options and the name of the tool which should be run.
There are basic tools such as \texttt{fft} which calculates the Fourier transform of an md-array,
as well as many commands for MRI-specific tasks such as whitening (\texttt{whiten}), coil compression (\texttt{cc}),
calibration of sensitivities (\texttt{ecalib}, \texttt{ncalib}), parallel imaging and compressed sensing (\texttt{pics}),
model-based reconstruction (\texttt{moba}), machine-learning reconstruction (\texttt{reconet}, \texttt{nlinvnet}),
and much more.

The tools act as building blocks and can be combined in a modular way to construct advanced
reconstruction pipelines for all kinds of MRI applications.
A complete reconstruction implemented with BART thus typically consists of a script that
calls various BART commands that operate on md-arrays.
For real-time reconstruction, it must be possible to split up, i.e. slice,
the multi-dimensional input space and process each subset of the data as it becomes available.
In other words, the reconstruction script has to be organized as a pipeline. To support this
in BART while preserving its modular character, the tools were enhanced to support 
streaming of slices of md-arrays.

\subsection{Slicing}

Suppose for $n\in\mathbb{N}$ that we want to split up an
$n$-dimensional array $X \in \mathbb{C}^{N_0 \times N_1 \times ... \times N_{n-1}}$.
In the following, the entries $X_{(i_0, ..., i_{n-1})}$ of $X$ are referenced using the multi-index
\[
(i_0, ..., i_{n-1}),~ i_j \in \mathbb{N}_0~\text{and}~i_j < N_j - 1.
\]

\begin{figure}[ht]
    \center
    \includegraphics[width=\textwidth]{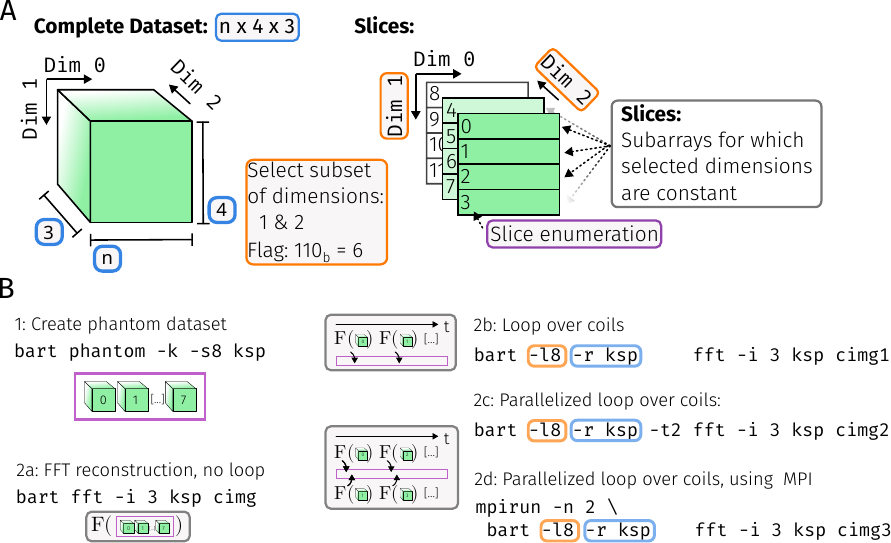}
    \caption{\textbf{A}: Schematic of how a three-dimensional array can be sliced.
    The cube on the left symbolizes the full three-dimensional array.
	By selecting a set of dimensions $\{2,1\} \subset \{2,1,0\}$,
	the array can be decomposed into a set of slices for which the
	index along dimension $0$ stays constant, as shown on the right.
    The set of selected axes can be efficiently encoded as a bitmask
    $6 = 1 \times 2^2 + 1 \times 2^1 + 0 \times 2^0$ (\cref{eq:bitmask}),
	and every slice is uniquely identified through a serial number.\\
\textbf{B}: Example command lines which demonstrate how slicing can be used in BART.
First, a k-space dataset with multiple channels is generated using command 1.
Commands 2a-d then calculate the FFT of it by using the \texttt{fft}-tool in different ways:
Command 2a processes the dataset in a single invocation (\textit{combined}).
Command 2b processes all channels serially.
This is specified using the loop option
\texttt{-l 8}, where $8=2^3$ represents the
coil sensitivity dimension,
and the reference file \texttt{-r ksp}
provides the loop limits.
Command 2c uses the \texttt{-t} (threads) option to
process all channels in parallel in two threads.
Note that 2a will typically use multiple threads as well,
but on a lower level within the FFT implementation itself.
Finally, 2d uses \texttt{mpirun} to process the dataset on two nodes \cite{mpi41}
or two GPUs.
}
    \label{fig:slicing}
\end{figure}

One method to subdivide this array is slicing \ie to select a set of axes, and
take those subsets of the array within which the indices of the selected axes stay constant.
This concept is illustrated in \cref{fig:slicing}A.
To give a precise description, if $A := \{0, 1, ..., n - 1\}$ is the set of all axes of the array,
then, for some $m < n,\, m\in\mathbb{N}$, a subset of $m$ axes $S \subset A$ is first selected for slicing.
Any given multi-index can then be decomposed into two multi-indices $a$ and $b$:
\begin{align}(i_{S(0)}, ..., i_{S(m - 1)})   &= (a_0, ..., a_{m - 1}),\\
    (i_{F(0)}, ..., i_{F(n - m - 1)}) &= (b_0, ..., b_{n - m - 1}),
\end{align}
where $F = A \setminus S$ is the set of free axes, and $S(k)$ and $F(k)$ denote the k-th smallest
entry of the set $S$ and $F$, respectively.
In other words, $X$ is split into a number of $n - m$ dimensional sub-arrays or \enquote{slices}
by fixing part of the multi-index in every slice.

Every slice is associated with one specific multi-index $a$. The set of all slices can be
enumerated by flattening the multi-index $a$ into a serial number, i.e. a single integer
\begin{align}
    c = \sum_{k = 0}^{m - 1} a_k \left(\prod_{l = 0}^{k - 1} N_{S(l)}\right). \label{eq:flat_index}
\end{align}
As common in BART, the selected set of dimensions $S$ used for this slicing operation can
be efficiently encoded as a single integer using a bitmask. This is done by interpreting the
pattern of selected/non-selected axes as a binary number
\begin{align}
    \text{bitmask}(S) = \sum_{i \in S} 2 ^ i. \label{eq:bitmask}
\end{align}

\subsection{Looping}

The idea of looping is to apply the functionality of a BART tool sequentially to
every slice, instead of applying it once to the whole array
(normal/\textit{combined} operation).
The invoked BART tool then does not \enquote{see} the entire array,
but instead operates on a single slice.
This can sometimes lead to semantic differences,
\eg when a full 3D reconstruction is numerically different to reconstruction
decoupled into slices.
In many cases, looping will lead to exactly the same result,
but modifies computing requirements and applicability:
\eg reducing memory requirements,
or enabling distribution across high-performance cluster
or multiple GPUs.

The looping functionality in BART is controlled
using a newly added set of options, referred to
as \textit{loop options}, which are passed
to the \texttt{bart} driver command.
More information on loop options can be found in the BART help,
accessible by calling \texttt{bart -h}, and the documentation folder
of the BART repository.
\cref{fig:slicing}B illustrates looping in a self-contained example.

\subsection{Streaming}

The goal of BART's streaming feature is to enable pipeline processing by
letting the tools send and receive slices of md-arrays during the computation.
This enables a chain of tools to simultaneously process different slices
as part of a single reconstruction pipeline.

BART will use streaming automatically  based on the
file name of the argument.
First, streaming is active if a hyphen is specified as file name,
which by convention is understood as a reference to standard input/output.
Using the vertical-bar or pipe operator \texttt{|}
available in common shells on
UNIX-derived systems \cite{Ritchie_Bell.Labs.Tech.J_1984},
a series of BART commands can be started in parallel,
such that the standard output of one command is connected to
the standard input of the next command.
Second, BART uses streaming
when a file name ends in \texttt{.fifo}.
\textit{Fifo}s (First-In-First-Out-Files / named pipes)
enable streaming of multiple in-/outputs at the same time, which
is needed to write complex reconstruction scripts with streaming support,
as described later in \cref{sec:streaming-apps}.

Generally, all BART tools directly support streaming.
By default, a tool will
wait (block) until all data in a stream has fully arrived.
This is essentially the same behavior that would also occur if no streams were used.
A more powerful method is to combine streaming and looping.
If a specified reference file is a stream, the stream flags are directly used as
loop flags and the BART tool is invoked on every slice as soon as it is received.
Thus, the combination of the loop functionality and
streaming allows
any BART tool to be used as part of a pipeline without blocking it.
Additionally, several tools have been made stream-aware,
to simplify certain operations such as regularization onto previous frames
in real-time applications of \texttt{nlinv}.

\FloatBarrier

\subsection{Applications}
\label{sec:streaming-apps}

We demonstrate the application of BART streams in three different exemplary scenarios:
a gridding reconstruction, a complete RT-NLINV reconstruction with preprocessing,
and a comprehensive reconstruction script accomodating various steps
for advanced RT-MRI.

\begin{figure}[ht]
    \center
    \includegraphics[width=.75\textwidth]{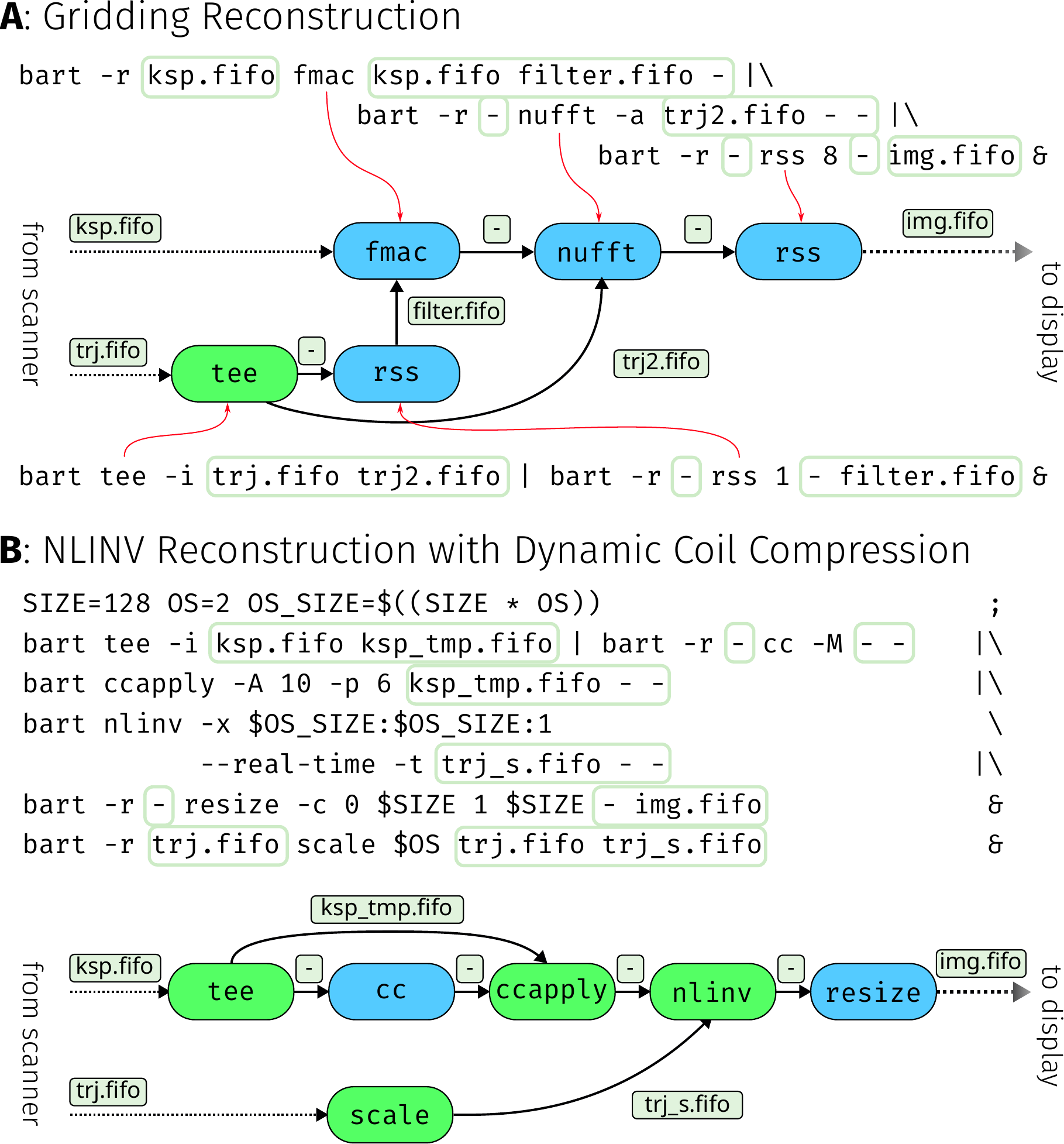}
    \caption{Part \textbf{A} shows a simple reconstruction pipeline using
a Ram-Lak filter, an adjoint NUFFT, and root sum-of-squares coil combination.
Shell code with which the pipeline can be created is shown along with
a representation of the computational graph.
Edges represent files, large colored nodes represent BART processes.
\texttt{ksp.fifo} and \texttt{trj.fifo} are named pipes,
which are expected to deliver incoming k-space data and trajectory from
the scanner or previous processing steps.
Similarly, \texttt{img.fifo} provides reconstructed images for further
processing and display.
Colors in the process nodes are used to highlight how tools can utilize
streams in different ways.
Blue commands use the loop options, while green commands are stream-aware.\\
Part \textbf{B} shows an example for an advanced reconstruction where
geometric coil compression and NLINV are
used for real-time reconstruction of radial data.
Incoming raw data is first split with the \texttt{tee} command and
forwarded to two different tools,
\texttt{cc} and \texttt{ccapply}.
The \texttt{cc} tool calculates the coil compression matrices, but
does not apply them.
Instead, they are forwarded to the \texttt{ccapply} tool, which applies
the received matrices to the incoming raw data and forwards the
compressed k-space to the \texttt{nlinv} reconstruction.
A subsequent \texttt{resize} command removes oversampling in the image domain.
}
    \label{fig:stream_cmds}
\end{figure}

To begin, reconstruction for RT-MRI using radial sampling can be
performed using the adjoint non-uniform FFT (NUFFT)
or \textit{gridding} \cite{OSullivan_IEEETrans.Med.Imag._1985}.
Here, the raw k-space data from the scanner is first multiplied
with the Ram-Lak filter \cite{Ramachandran_PNAS_1971},
to compensate for non-uniform sampling density of radial sampling.
Afterwards, the adjoint NUFFT is applied to obtain coil images,
which are combined using root sum-of-squares.
Compared to iterative reconstructions, this scheme is relatively simple and
can run in real-time even without GPU acceleration.
\cref{fig:stream_cmds}A shows an RT-MRI pipeline implementation
of this reconstruction using BART streams.

\cref{fig:stream_cmds}B shows a more sophisticated reconstruction
using RT-NLINV, which however still requires only a few BART commands.
Here, we also employ a new dynamic variant of
geometric coil compression \cite{Zhang_Magn.Reson.Med._2013}
which is adapted to RT-MRI by performing alignment of
the compression matrices along the time dimension \cite{Schaten__2024}.
Using this method requires incoming data to be split up, processed, and
then recombined in real-time.

As a final example, a complete state-of-the-art
reconstruction pipeline for
radial real-time MRI was created.
It takes the form of a BASH script
which launches different reconstruction pipelines, based either on
the gridding-based reconstruction (\cref{fig:stream_cmds}A)
combined with a sliding window strategy
\cite{Zhang_J.Magn.Reson.Imaging_2010}, or on RT-NLINV.
Additional preprocessing steps such as coil compression
and gradient delay correction based on RING
\cite{Rosenzweig_Magn.Reson.Med._2018a}
and post processing steps such as a median filter
or non-local means filter \cite{Buades_2005CVPR_2005}
can be selected by the user.
A graph representation of one reconstruction pipeline
generated by this script is available in supporting material
\ref{sec:graph}.

\subsection{MRI Scans}
\label{sec:scans}
We show exemplary results of
the real-time reconstruction script on
cardiac scans of a healthy adult volunteer,
who was examined with approval by the local ethics board
and after giving written informed consent.
As this work mainly focusses on technical implementation,
our findings are largely independent of subject variability,
and should generalize well.
Measurements were done with a Magnetom Vida 3T
(Siemens Healthineers, Erlangen, Germany)
using a radial FLASH \cite{Haase_J.Magn.Reson._1986} sequence
with random RF-spoiling \cite{Roeloffs_Magn.Reson.Med._2016}
and a turn-based pattern with 5 turns and
13 spokes per turn \cite{Zhang_J.Magn.Reson.Imaging_2010}.
Further parameters are given in \cref{tab:params}.
In addition, phantom scans were performed
to measure
end-to-end latency as described in \cite{Schaten__2025},
\cf supporting material \cref{fig:latency_method}.

All images shown are reconstructed
using the real-time reconstruction
script introduced in \cref{sec:streaming-apps}.
We compare images from different quality settings:
\begin{itemize}
    \item \textit{Fast}: Gridding-based reconstruction,
         with static coil compression to four channels.
    \item \textit{Good}: Gradient delay correction,
        coil compression to eight channels,
        RT-NLINV-reconstruction and five-frame median-filtering
    \item \textit{High}: Gradient delay correction,
        aligned coil compression to eight channels,
        RT-NLINV reconstruction with more iterations,
        median filtering and
        additional non-local means filtering
        \cite{Buades_2005CVPR_2005}.
\end{itemize}
\textit{Good} and \textit{High} quality reconstructions were
accelerated using an Nvidia H100 GPU (80 GB HBM3, SXM)
on a system with an AMD EPYC 9334 CPU.
\textit{Fast} quality reconstruction is done without GPU acceleration.

Additionally, we assess the impact of
aligned dynamic coil compression
as described in \cref{sec:streaming-apps}.
This is based on the \textit{good}
quality setting, but omitting any post-processing steps.
We compare the following choices for coil compression (CC):
Static CC using data from the first frame,
static CC using data from all frames (non-causal),
dynamic CC on every frame separately, and
aligned dynamic CC.

Reconstructions are run on a server in the neighbouring building,
connected to the acquisition system via 1 Gigabit Ethernet.
Data is exchanged with the MRI scanner using  BART streams 
with a custom data import/export software
that is integrated
into the vendor software (ICE) running on the scanner,
which sends out every
k-space line over the network as soon as it was acquired.
Conversely, every received image is forwarded
to ICE for display as soon as it is received.

\begin{table}[]
\centering
\caption{Parameters of the RF-spoiled radial FLASH MRI measurements.}
\label{tab:params}
\begin{tabular}{@{}lll@{}}
\toprule
    Parameter       & Value         \\ \midrule
    ADC Samples (2 $\times$ oversampling)    & 256 \\
    Spokes / frame	& 13 \\
    Matrix size & 192 \\
    Flip angle      & \qty{10}{\degree} \\
    TR              & \qty{2.1}{ms} \\
    TE              & \qty{1.25}{ms}\\
    FOV             & \qty{256}{mm} \\
    Bandwidth       & \qty{1300}{\hertz\per\pixel} \\
    Slice thickness & \qty{5}{mm}   \\ \bottomrule
\end{tabular}
\end{table}

\subsection{Latency and Performance Measurements}
\label{sec:methods_latency}

    Computational performance of the new BART features was evaluated.
With only a single thread on one compute node,
looping will typically be slower than combined operation
due to overhead such as repeated initialization.
We first quantify this overhead
imposed by streaming/looping.
Furthermore, we measure the
peak amount of memory which was simultaneously used by BART
in the computer's main memory (resident set size).

First, an md-array of complex random numbers
\texttt{x} $\in\mathbb{C}^{n_r \times n_p \times n_s}$ is created,
representing raw data from a Cartesian multi-slice MRI measurement
for different matrix sizes $n_r,\, n_p$
with oversampling in readout direction $n_r = 2n_p$
and a varying number of slices $n_s$.
An inverse 2D-FFT of every slice is then performed in three different ways (\cf \cref{fig:slicing}).
First, with a regular BART command, resulting in combined processing of the data:
\begin{lstlisting}
bart fft -i 3 x out
\end{lstlisting}

Second, the 2D-FFT is sequentially performed on every slice using the loop options:
\begin{lstlisting}
bart -l4 -r x fft -i 3 x out
\end{lstlisting}

Third, data is read by the looped \texttt{fft} command
from a previously created text file using the stream protocol:
\begin{lstlisting}
bart -l4 -r x copy x - > stream.txt
bart -l4 -r - fft -i 3 - out < stream.txt
\end{lstlisting}

In a second experiment, we measure the speed-up for a simple reconstruction
pipeline where streaming and looping enable parallelization of consecutive
reconstruction steps.
A customizable delay is added to the input stream,
simulating per-slice acquisition time.
The output of the \texttt{copy}-tool is then
forwarded to the \texttt{fft}-tool directly via a pipe,
causing the processes to run in parallel:
\begin{lstlisting}
bart -l4 -r x copy --delay <delay> x - |\
bart -l4 -r - fft -i 3 - out
\end{lstlisting}

This is compared to the static variant:
\begin{lstlisting}
bart -l4 -r x copy --delay <delay> x - |\
bart fft -i 3 - out
\end{lstlisting}
where the reconstruction can only begin after the \texttt{copy}-process has processed all slices.
The experiment is repeated for different delays between \qty{0}{ms} and \qty{130}{ms}.
Wall-clock run time is measured using the \texttt{date} program from GNU Coreutils.
The experiments were repeated 25 times to measure standard deviation across runs,
and performed on a standard workstation using an
Intel Xeon W-2123 CPU running at \qty{3.60}{GHz}.

To assess latency for real-time MRI, phantom measurements were performed.
We use the real-time reconstruction script
(\cref{sec:scans}, set to \textit{good}),
and additionally
run the experiment without median filter to
assess the impact of this post-processing step on overall latency.
Apart from acquisition and the reconstruction, several other factors such as network transfer times,
filters used in post processing, and image display are expected to contribute to the total latency.
Therefore, we determined several different latencies:
\begin{itemize}
    \item Latency in BART is calculated as time difference between
arrival of the last spoke and transmission of the reconstructed frame.
Preprocessing, reconstruction and post processing were also measured individually.
Timestamps are recorded using the \texttt{gettimeofday} function from glibc.
\item Latency BART + Network as seen from the software running on the scanner
is measured by taking timestamps when the last spoke is sent over the network
and after the corresponding frame has been received. This also includes data transfer
times and network delays.
\item End-to-end latency, \ie the time between an event happening in the scanner
until it is shown on the MRI screen, is measured using an experimental setup
as described in \cite{Schaten__2025}
, correlating movements in a phantom
with movement on the screen using a video recording.
Additional information is available in supporting material \ref{sec:e2e_latency_method}.

\end{itemize}

Because the MRI acquisition starts as soon as possible
without waiting for the BART reconstruction pipeline to be initialized,
the latency decreases towards a steady state,
which is reported here as \textit{steady state latency}.

\section{Results}

\subsection{Performance}
\label{sec:results_performance}

\subsubsection{Numerical experiments}

\begin{figure}[ht]
    \center
    \includegraphics[width=\textwidth]{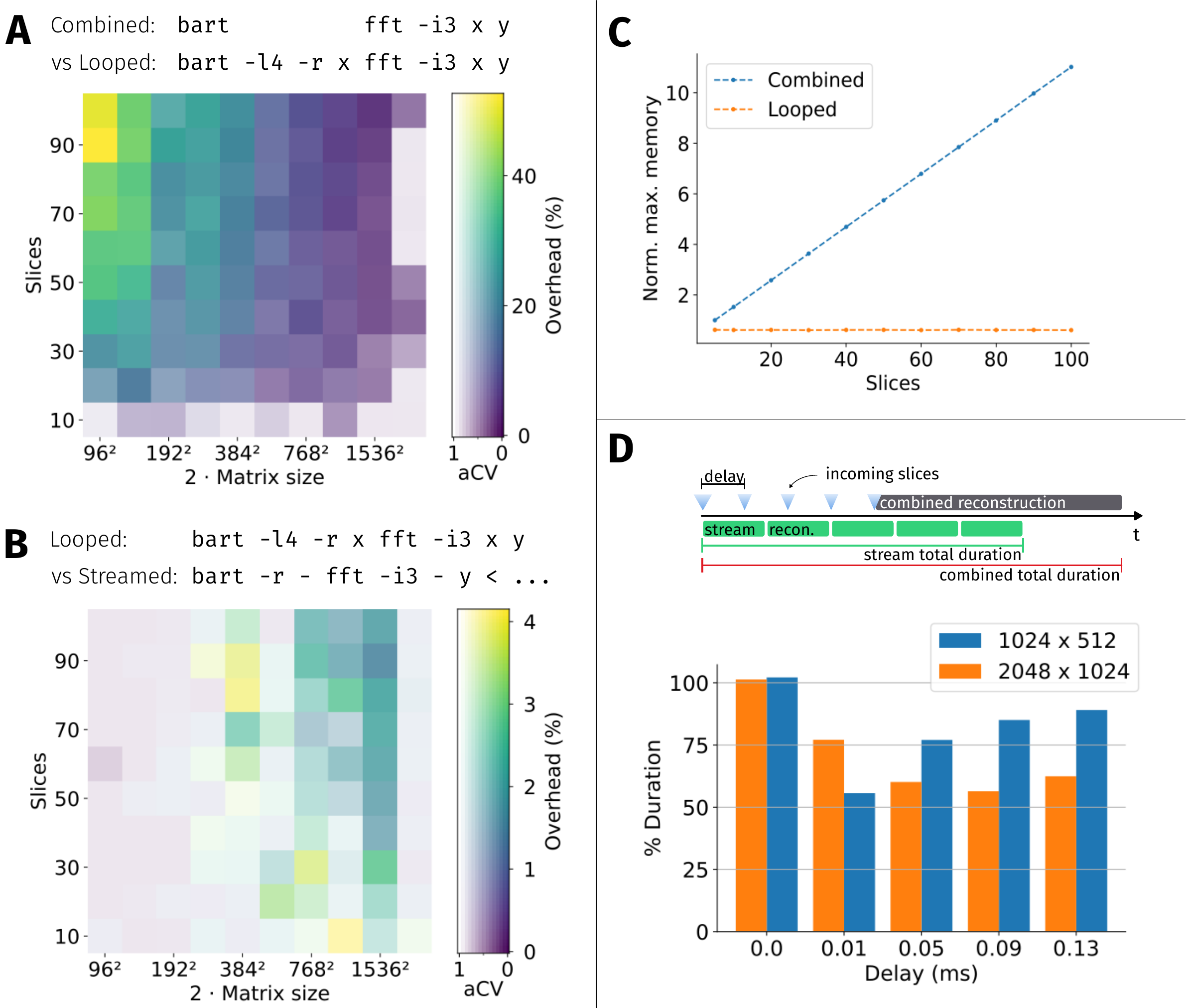}
    \caption{The performance of BART looping and streaming is shown here
        by visualizing the runtime of an inverse FFT
        on arrays of varying size.\\
        Part \textbf{A} shows a heat map of additional time required when
        performing a looped FFT relative to the normal/combined FFT,
        for different problem sizes.
        The opacity/lightness is based on an adjusted coefficient of variation (aCV):
        Fully saturated colors represent a distinct overhead,
        while light colors indicate that the average run-to-run difference
        is at least as big as the calculated overhead.\\
        Part \textbf{B} compares the overhead of streaming relative to looping, similar to part A.\\
        Part \textbf{C} visualizes the memory advantage of looped versus
        combined operation for a matrix size of 512 x 256 with varying
        number of slices. Memory usage is normalized to the amount required
        for a single slice, \ie array size 512 x 256 x 1.\\
        Part \textbf{D} shows the speed-up due to reduced waiting times
        in pipeline operation, \ie when the process producing the data
        (acquisition) and the reconstruction can be parallelized.}
    \label{fig:bench}
\end{figure}

We first present results on the performance of the two main additions
to the BART software, streaming and looping.
As described in \cref{sec:methods_latency}, looping and streaming can under some conditions
create computational overhead.
\cref{fig:bench}A shows the overhead when performing
a looped FFT compared to normal FFT. For small arrays with many iterations / slices,
there is a large relative overhead.
E.g., the FFT of a 96x48x100 array takes 50\% or \qty{10(1)}{ms}
(mean with standard deviation in parentheses)
longer when looping over the last dimension.
FFT of a 2048x1024x100 array takes 2\% or \qty{400(150)}{ms} longer.
For few slices or large problem sizes, the differences between looping and
combined operation are small relative to the variation over repetitions.

In \cref{fig:bench}B, we repeat this analysis for the streaming protocol ,
comparing the extra time required when performing looping with a streamed input
as opposed to looping alone.
The maximum overhead is several times smaller than in the previous analysis.
Furthermore, the overhead induced by the stream protocol is smaller than
the average run-to-run variation in most cases.
Also, there is no clear trend towards
any greater or smaller overhead depending on the matrix size.

\cref{fig:bench}C, illustrates how, with looping, peak memory usage
is independent of the number of slices while it grows linearly
with normal \ie combined operation.
Furthermore, combined streaming and looping enables parallelization of
successive data processing steps, and thus can speed up computation.
This is illustrated in \cref{fig:bench}D.
Depending on matrix size and delay (slice acquisition time),
total elapsed time was reduced down to a minimum of 56\% of the reference.

\subsubsection{End-to-End Latency}

\begin{figure}[ht]
\center
\includegraphics[width=\textwidth]{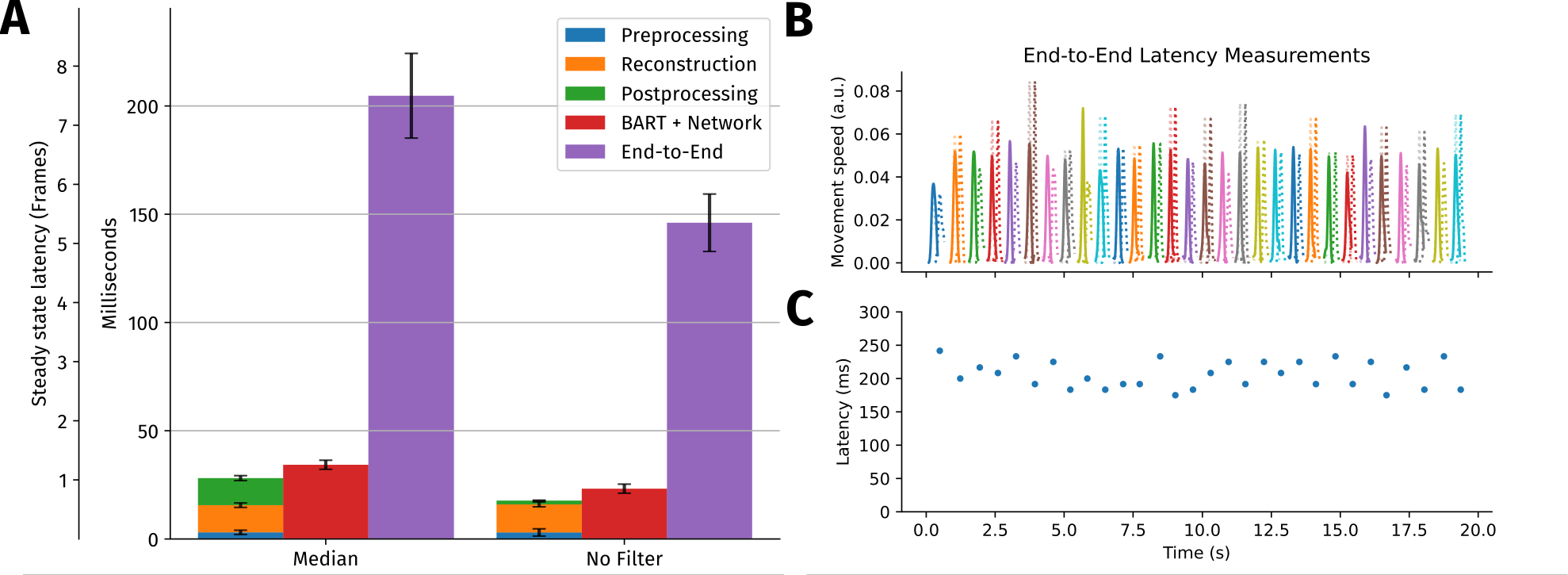}
\caption{Latency measurements for RT-MRI.
Part \textbf{A} visualizes results from three different latency measurement methods described previously.
Latency determined within BART is further separated into three different steps:
preprocessing, reconstruction and post processing.
The times are shown for two measurements,
with and without 
median filter in the post processing chain.
Error bars indicate standard deviation over the course of the time series.
An initially higher end-to-end latency in the first few seconds is excluded
for calculating the steady-state end-to-end latency shown here.
\\
Part \textbf{B} on the right highlight details
of the end-to-end latency measurement.
Every colored line corresponds to a movement in the end-to-end
latency measurement setup.
Solid lines are calculated from the optically observed movement, while dashed
lines correspond to the movement observed in the reconstructed images shown
on the MRI console. The optimally shifted MRI movement curves are shown with
transparent dashed lines.
In \textbf{C}, the resulting time series of end-to-end latency measurements is shown.}
\label{fig:latency}
\end{figure}

The latency of our RT-MRI setup was measured
using different methods (\cref{sec:methods_latency}).
The measurement was repeated three times, with almost identical outcome.
One instance of these measurements is shown in \cref{fig:latency}.
The BART-internal timing measurements reveal the amounts of time
required for preprocessing, reconstruction and post processing,
which are \qty{3(2)}{ms}, \qty{13(2)}{\ms} and \qty{13(1)}{ms}, respectively.
Relative to the acquisition time of a single frame, which is \(13\cdot2.1\,\)\si{ms}\(=27.3\,\)\si{ms},
all steps are relatively fast.
Without the median filter post processing time is \qty{2(1)}{ms}.

The latency measured from the perspective of the software running on the scanner
matches the total latency measured within BART with a difference of 
approximately \qty{5}{ms} corresponding to the network transfer.
The end-to-end latency of \qty{205(20)}{ms} is several times larger 
than the other two latency measurements with only \qty{28(2)}{ms} from
the BART reconstruction.
This discrepancy between end-to-end latency compared to the other measurements
has multiple sources.
For instance, the acquisition time itself contributes to the end-to-end latency.
Furthermore, time-dependent filters such as the median filter do not affect
the signal processing time too much, but have a large impact on the end-to-end latency.
Without median filter, an end-to-end latency of \qty{146(13)}{ms} is achieved.
Several other possible sources are treated in the discussion.

\subsection{Real-Time Applications}

\begin{figure}[ht]
    \center
    \includegraphics[width=\textwidth]{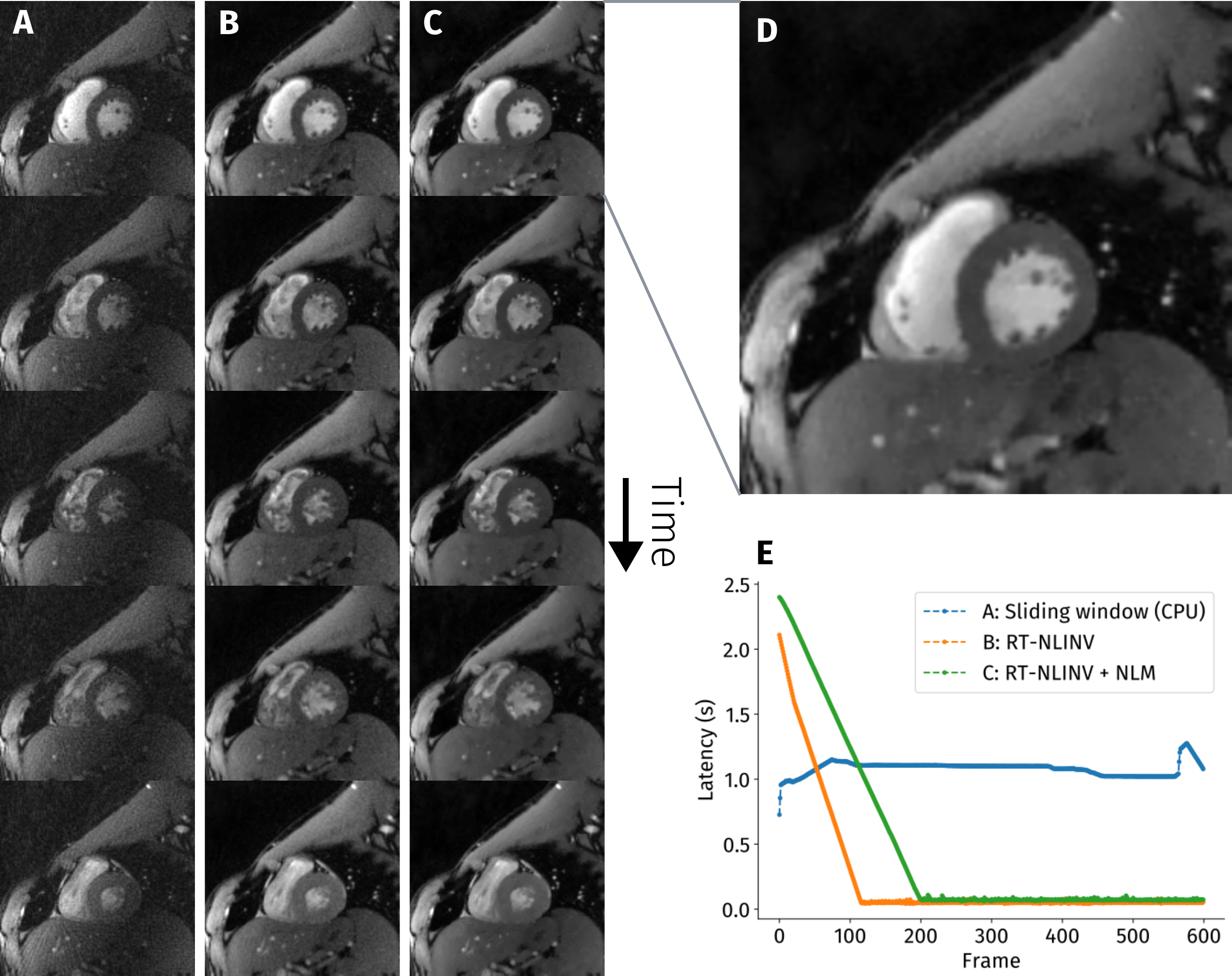}
    \caption{Five representative frames covering one heartbeat
        from a real-time acquisition
    	of a cardiac short-axis view, reconstructed with
    different quality settings for the real-time reconstruction
        script (\cref{sec:scans}).
    Column \textbf{A} shows 
    images reconstructed with
    \textit{fast},
    column \textbf{B} \textit{good} and
    column \textbf{C} \textit{high} quality settings.
Part \textbf{D} provides an enlarged view of the \textit{high} quality setting.
The graph \textbf{E} shows the numerically determined latency
of all configurations.
}
    \label{fig:algorithms}
\end{figure}

In our experimental setup, the simple reconstruction based on the adjoint NUFFT
can be run in real-time on standard hardware  without use of a GPU as demonstrated 
by the latency graph shown in \cref{fig:algorithms}E.
Iterative reconstruction methods such as RT-NLINV can be used to achieve high image quality,
which can be further improved through post processing, as seen in \cref{fig:algorithms}B-D.
The initially higher, but then quickly decreasing latency in these cases is
probably due to initialization of the GPU and precomputing steps.
A comparison between non-causal reconstruction and real-time capabable reconstruction
can be found in the literature \cite{Uecker_Magn.Reson.Med._2010}.
Videos of the reconstruction are available in
supporting material \ref{sec:rtreco_vid}.

\begin{figure}[ht]
    \center
    \includegraphics[width=\textwidth]{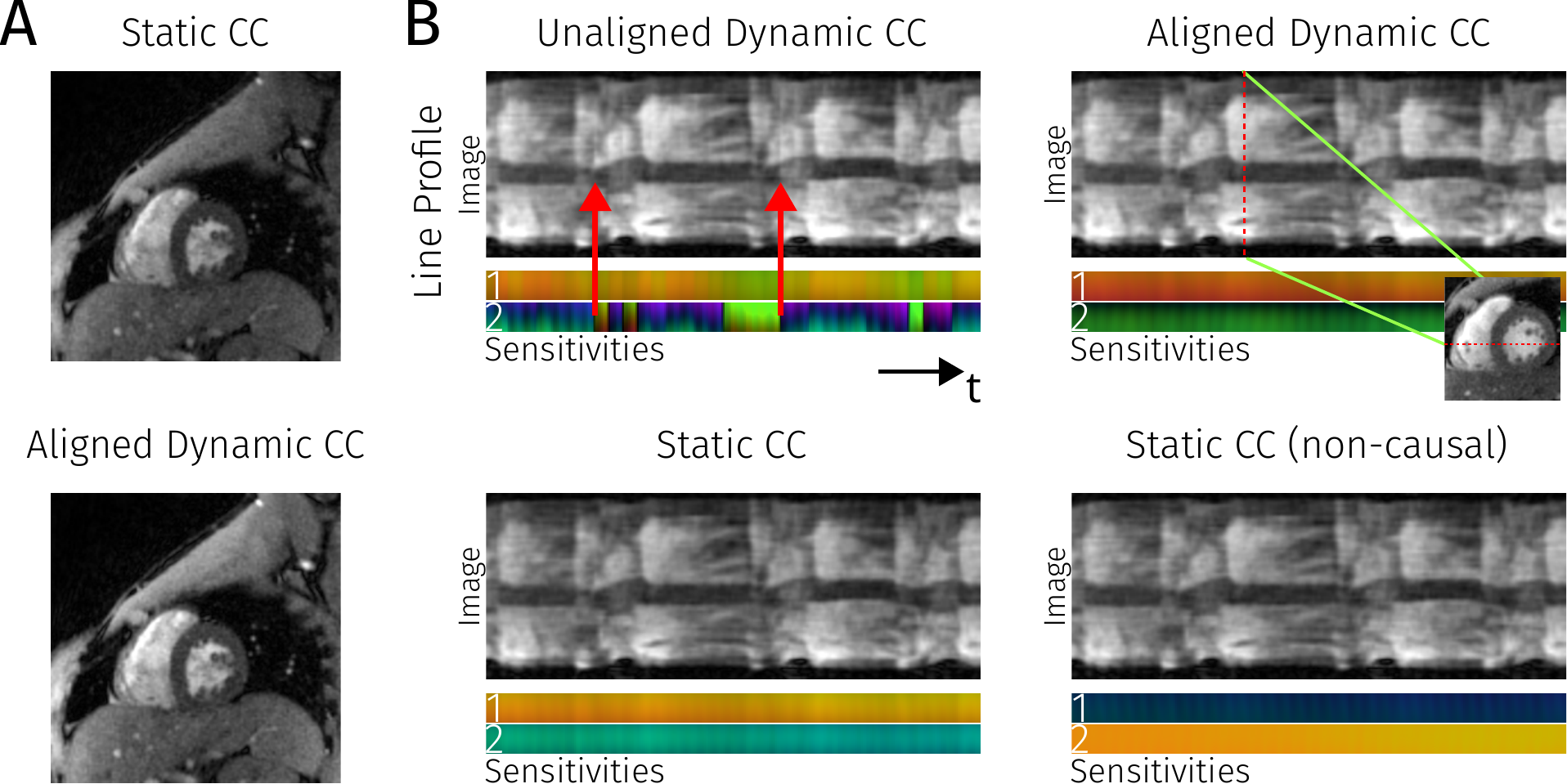}
    \caption{Part \textbf{A} compares two frames from a RT-MRI of the heart
        with different strategies for coil compression,
        either re-using the compressing matrix or
        recomputing and aligning it for every frame.
	The red arrows highlight abrupt changes in image magnitude which correlate with the discontinuities in the virtual coil profiles.
        Part \textbf{B} shows a time series from a line through the heart,
        with the line being shown in the small inset image.
        The bottom row shows the time series for the first two
        virtual coil profiles, the top row shows the actual images.
        On the top left, every frame was independently compressed,
        on the top right, compression matrices were aligned
        using geometric coil compression.
        For comparison, line profiles for
            coil compression done on the first frame/complete dataset
        are shown in the bottom left/right.
}
    \label{fig:gcc}
\end{figure}

\cref{fig:gcc} 
    compares static and dynamic coil compression and demonstrate
the impact of aligned dynamic coil compression
as described in \cref{sec:streaming-apps}.
In \cref{fig:gcc}A, it can be seen that noise is increased after a change of slice position,
as the compression matrix is not adapted to the changed coil profiles in the new slice.
Dynamic coil compression
leads to problems as shown in \cref{fig:gcc}B:
The coil profiles show distinct jumps over their time course
and this then also affects the reconstructed images.
These discontinuities disappear when using 
aligned dynamic coil compression.
For comparison, we also show coil and image profiles
from reconstruction with static coil compression,
which exhibit no
discontinuities/brightness jumps.

\section{Discussion}

Streaming functionality was added to the BART reconstruction software
as new feature. The implementation achieves high modularity and interoperability
for real-time reconstruction pipelines by relying on standard operating system utilities.
Based on a set of new options, all tools included in the toolbox can be
used in real-time reconstruction pipelines.

The streaming extension was tested and benchmarked in numerical experiments
and using phantom and in-vivo measurements.
As it is now possible to parallelize subsequent
reconstruction steps in a pipeline, a speed-up can be achieved
for conventional reconstruction pipelines simply by chaining BART commands
.
Under certain circumstances, the overhead associated with streaming and looping can
have a negative effect on performance, but in all tested conditions overhead 
was either small in absolute units, small compared to the total computation time,
or small relative to the run-to-run variations.  In most practical applications,
the overhead  should not be relevant or even be unnoticeable.

Naturally, there is some overlap between BART and other reconstruction
frameworks, such as the Gadgetron \cite{Hansen_Magn.Reson.Med._2013} and also with
vendor pipelines. Nevertheless, BART is not meant to replace such other
frameworks, but to provide a set of versatile tools that can be used to
solve specific problems in a variety of scenarios. In fact, it is very common
to only use specific  BART tools as part of other projects and this is
facilitated by BART's open and modular design. So far, this was exploited 
mostly for offline reconstruction. By providing a low-latency path for
continuous data exchange, 
BART can now also be integrated into real-time processing pipelines.

To demonstrate the flexibility of the new streaming framework, we showed examples
that extend from simple NUFFT-based reconstruction which can be used even on
desktop systems to advanced iterative reconstruction that achieves image quality
and makes use of GPU acceleration.  The advantages of dynamic coil compression
was shown, demonstrating how
advanced functionality can be implemented in a modular pipeline. We provide a
versatile reconstruction script for real-time as part of the BART toolbox that
includes all the functionality described in this work.
New BART functionality such as removal of phase pole artefacts \cite{Blumenthal__2025}
were also already integrated and shown to be beneficial in preliminary experiments.

We furthermore measured the latency of our
interactive RT-MRI setup for a full reconstruction pipeline.
Here, we found that the part of the latency corresponding
to the BART pipeline is
only a small fraction of the overall end-to-end latency.
Processing time in BART, including network transfer,
is below \qty{30}{ms}, which is acceptable even for
demanding application such as cardiac interventional MRI.
End-to-End latency is considerably higher with \qty{200}{ms},
which is in part explained by the response of the median filter
used in post processing, and the duration of the
data acquisition itself. As shown previously \cite{Frahm_TOMIJ_2014}
the temporal regularization does not add substantial latency. 
Another source of end-to-end latency is image processing
in the vendor software, which includes \eg distortion correction
and displaying of the images.

For integration with the MRI scanner, we used a custom,
vendor-specific software for data import/export.
However, we also already performed initial experiments using scanner integration
based on MRD streams, using the BART \texttt{ismrmrd} tool to interface
at the start/end of the reconstruction pipeline.
With vendors increasingly providing open interfaces, this should
simplify the integration of BART into custom pipelines.

\section{Conclusion}

The streaming extension to the BART framework enables
efficient reconstruction for RT-MRI applications with low latency, while
retaining the full modularity of the BART framework.

\FloatBarrier

\section*{Conflict of Interest}
The authors declare no competing interests.

\section*{Data Availability Statement}
In the spirit of reproducible research, the code to reproduce the results
of this paper is available at
\url{https://gitlab.tugraz.at/ibi/mrirecon/papers/bart-streams} (version v0.2).
All reconstructions have been performed with BART, available at
\url{https://github.com/mrirecon/bart}
and \url{https://codeberg.org/mrirecon/bart}.
The data used in this study is available at Zenodo \doi{10.5281/zenodo.17671124}.

The authors are committed to supporting BART for at least another ten
years. Long-term reproducibility is ensured by regular
automated testing. A mailing list is available for questions and problems
can be reported via a publicly available contacts. Code contributions are welcomed.

\section*{Acknowledgements}
\printfunding

The authors thank the ISMRM Reproducible Research Study Group
for conducting a code review of the code (Version v0.1) supplied
in the Data Availability Statement.
The scope of the code review covered only the code’s ease of download,
quality of documentation, and ability to run,
but did not consider scientific accuracy or code efficiency.

\printbibliography

\clearpage
\appendix
\section*{Supporting Information}
List of supporting material, including captions.

\subsection{Real-Time MRI Reconstruction Pipeline Graph}
    \label{sec:graph}

\begin{figure}[ht]
    \center
    \caption{
State-of-the-art reconstruction pipeline for real-time MRI.
Different sections of the reconstruction pipeline are highlighted, and
characteristic parts are also shown in a zoomed-in, abbreviated version next to the full graph.
Round nodes represent BART invocations while edges between tool nodes represent pipes.
The invocation of the real-time reconstruction script corresponding to this graph is shown at the bottom,
along with an explanation of command line options.
The figure is based on an auto-generated graph, which was created by tracing every BART tool invocation
and translating this trace into a graph description language,
and was rendered with \texttt{dot} \cite{Gansner_Software_2000}.
}
    \label{fig:graph}
\end{figure}

\subsection{End-To-End Latency Measurement}
    \label{sec:e2e_latency_method}
\begin{figure}[ht]
    \center
    \caption{Part \textbf{A}: Schematic overview of the end-to-end latency measurement experiment setup.
A water-filled test-tube (blue) inside the MRI scanner is pulled up from the control room using a long string (yellow).
Images from the real-time sequence are shown on the MRI console screen.
A smartphone camera records movement on the MRI screen and movement of a paper tag (gray) attached to the string.
A single frame from the resulting video is shown on the top right.\\
Part \textbf{B} illustrates how the resulting video is processed:
The video is segmented, and the obtained object velocity curves are cross-correlated
to obtain the delay between both movements.}
    \label{fig:latency_method}
\end{figure}

\subsection{Real-Time MRI Reconstruction Videos}
    \label{sec:rtreco_vid}
Corresponding video for Figure 5, covering three heartbeats,
starting at the same time as the timeseries shown in Figure 5.
The ordering matches the original figure, that is, images were reconstructed
using (from left to right) the fast, good, and high-quality settings of the
rtreco.sh reconstruction script.
\setcounter{figure}{3}

\end{document}